%% file: MunariLevoratoZorzi_CoopInCSMA_part2.tex
\documentclass[a4paper, draftcls, onecolumn, 12pt]{IEEEtran}
\usepackage[noadjust]{cite}
\usepackage{latexsym}
\usepackage[latin1]{inputenc}
\usepackage[T1]{fontenc}
\usepackage[dvips]{graphicx}
\usepackage{verbatim}
\usepackage{amsmath}
\usepackage{amsfonts}
\usepackage{amssymb}
\usepackage{exscale}
\usepackage{latexsym}
\usepackage{color}
\usepackage{bm}
\usepackage{stfloats}
\usepackage[nolists, nomarkers]{endfloat}
\usepackage[letterpaper, textwidth=6.5in, textheight=9in]{geometry}

\newcommand{\figw}{0.98\columnwidth}

\author{\large{Andrea Munari$^{\ddag}$,~\IEEEmembership{Member,~IEEE}, \\Marco Levorato$^{*}$,~\IEEEmembership{Member,~IEEE}, Michele Zorzi$^{\dag}$,~\IEEEmembership{Fellow,~IEEE}}\\
\normalsize $\ddag$ German Aerospace Center (DLR), Inst. of Communications and Navigation, Oberpfaffenhofen, 82234 Weßling, Germany. \\
\normalsize $*$ Dept.\ of Electrical Engineering, Stanford University, Stanford, CA 94305 USA. \\
\normalsize $\dag$ Dept.\ of Information Engineering, University of Padova, via Gradenigo, 35131 Padova, Italy. \\
\normalsize e-mail: andrea.munari@dlr.de, levorato@stanford.edu,
zorzi@dei.unipd.it.
\thanks{Andrea Munari was with the Dept.\ of Information Engineering, University of Padova, via Gradenigo, 35131 Padova, Italy.}}

\date{}

\title{Cooperation in Carrier Sense Based Wireless Ad Hoc Networks - Part II: Proactive Schemes}

\begin{document}

\setcounter{page}{0}
\maketitle
\thispagestyle{empty}

\vspace{-5mm}
\begin{abstract}
This work is the second of a two-part series of papers on the effectiveness of cooperative techniques in non-centralized carrier sense-based
ad hoc wireless networks. While Part~I extensively discussed \emph{reactive} cooperation, characterized by relayed transmissions triggered by
failure events at the intended receiver, Part II investigates in depth \emph{proactive} solutions, in which the source of a packet exploits channel state information to preemptively coordinate with relays in order to achieve the optimal overall rate to the destination. In particular, this work shows by means of both analysis and simulation that the performance of reactive cooperation is reduced by the intrinsic nature of the considered medium access policy, which biases the distribution of the available relays, locating them in unfavorable positions for rate optimization. Moreover, the highly dynamic nature of interference that characterizes non-infrastructured ad hoc networks is proved to hamper the efficacy and the reliability of preemptively allocated cooperative links, as unpredicted births and deaths of surrounding transmissions may force relays to abort their support and/or change the maximum achievable rate at the intended receiver. As a general conclusion, our work extensively suggests that CSMA-based link layers are not apt to effectively support cooperative strategies in large-scale non-centralized ad hoc networks.
\end{abstract}


\newpage
\input{introduction}

\input{csma}
\input{preAllocDescription}
\input{preAllocAnalysis}
\input{preAllocSims}
\input{conclusions}

\bibliographystyle{IEEEtran}
\bibliography{IEEEabrv,TWbib}

\end{document}

%% file: introduction.tex
\section{Introduction} \label{sec:introduction}

Cooperative communications have attracted an ever growing interest in the research community since the seminal works of Laneman and Sendonaris \cite{Laneman-dec04,Sendonaris-nov03,Sendonaris-nov03-2}. In this perspective, a great deal of effort has concentrated on the definition of novel cooperative paradigms and on the analysis of their performance from a theoretical angle \cite{Lai06,capth,cover_relay}, whereas somewhat less attention has been devoted so far to the issues that may arise when cooperation has to be applied in large networks and to the mutual influence between relaying strategies and link layers. Both these aspects, however, are of pivotal importance in view of the practical implementation of cooperation, as several problems that cannot be accounted for in simple scenarios may come into play and substantially alter the theoretical performance gains.

Under this line of reasoning, this two-part work is dedicated to the investigation of the impact that Carrier Sense Multiple Access (CSMA) has on cooperative relaying strategies in large scale non-centralized ad hoc networks. In the companion paper \cite{part1}, we focused on \emph{reactive} cooperation, characterized by a source node transmitting at a fixed information bitrate and by a relay terminal offering support by sending redundancy to the destination in the event of a failure over the direct link. This class of approaches implements a distributed Hybrid Automatic Repeat reQuest (HARQ) policy, and does not require any prior knowledge of the channels connecting the terminals involved in the data transfer, significantly easing the design and the implementation of cooperative solutions in the completely distributed scenarios that are the focus of our study. The work in \cite{part1} shows, by means of both analysis and simulation, that the effectiveness of reactive schemes is significantly hampered by the presence of several concurrent links in the network, as well as by the intrinsic nature of the contention mechanism. A first and key detrimental effect is represented by the strong spatial and temporal correlation on the interference level perceived by nodes close to each other, e.g., a source, a relay and a destination, which reduces the channel diversity gain that underpins all the cooperative benefits. Furthermore, CSMA is shown to hinder relaying by biasing the spatial distribution of the available cooperators, reducing the probability that one of them lies in the region that would maximize the throughput gains. 

These results allow to draw some important conclusions on the issues that beset a specific class of relaying solutions in realistic networking environments. Recently, however, several works have studied from a theoretical perspective a different kind of cooperative policies that take advantage of Channel State Information (CSI) to tune transmission rates, reporting significant gains over plain ARQ in ad hoc networks \cite{Lin06,Panwar07,Lichte09}. The basic idea of these schemes, which we refer to as \emph{proactive} cooperation,\footnote{The partition of cooperative policies in proactive and reactive was first introduced in \cite{Shan09}.} is to let a source choose whether to directly transfer its payload to the intended destination or to split the communication in two phases, delivering first its data to a relay and then letting this node send redundancy to the destination so as to exploit spatial diversity. The decision is made so as to minimize the overall transmission time, i.e., so as to maximize the per-link throughput.

In view of this, to further extend the conclusions of \cite{part1} and to gather a full understanding of the issues that have to be faced in realistic scenarios, this paper extensively investigates the effectiveness of proactive cooperation in non-infrastructured networks based on carrier sense. We start our study by showing how proactive policies enable significant gains with respect to non-cooperative counterparts in simple environments with few nodes and idealized medium access control. By means of both analysis and simulation, however, we also prove that these improvements dramatically shrink in practical distributed ad hoc networks, even under the assumption of perfect CSI knowledge in the coordination phase. As in the reactive case \cite{part1}, the disagreement between theoretical limits and actually achievable performance gains is partially due to the bias in the spatial distribution of the available relays induced by the contention mechanism. Such an effect, indeed, is intrinsic to carrier sense, and does not depend on the type of cooperative mechanism in place.

In addition, we extensively discuss how the effectiveness of proactive relaying depends on the stability of the CSI used to compute the transmission rates. From this viewpoint, indeed, a change in the boundary conditions that drive the cooperative policy in the preliminary coordination phase may unexpectedly lower the sustainable information bitrate at the destination, or may force a relay-elect to refrain from transmitting due to carrier sense as an effect of newly determined channel conditions. In this perspective, the highly dynamic nature of interference that characterizes non-infrastructured large-scale ad hoc networks is shown to jeopardize the efficacy of cooperative links.

Finally, this work also considers the impact of several practical issues that are typically neglected in the literature, such as synchronization and the hidden terminal problem, on proactive cooperative protocols.

The rest of the paper is organized as follows. Sections~\ref{sec:csma} and~\ref{sec:preAllocDescription} describe the carrier sense-based contention mechanism and the cooperative approaches under investigation, respectively. Section~\ref{sec:preAllocAnalysis} presents an analytical investigation
of the theoretical performance of the protocol, as well as the derivation of the spatial distribution of available relays. Section~\ref{sec:preAllocSims} extensively discusses the performance of the considered cooperative strategy in a detailed networking scenario, while Section~\ref{sec:conclusions} draws the conclusions of the paper.

%% file: csma.tex
\section{Carrier Sense Based Multiple Access and System Model} \label{sec:csma}

For the sake of clarity and self-containment, we report in this section, as done in \cite{part1}, a brief description of the key features of CSMA as well as an introduction of the system model considered throughout our work. 

Carrier sense is a fundamental networking mechanism, widely implemented in present day wireless systems to coordinate the simultaneous access to the medium of several nodes. In particular, thanks to its simplicity and to its completely distributed nature, such a control policy has become the paradigm for link layers in non-centralized ad hoc networks, which are the focus of our study. According to Carrier Sense Multiple Access (CSMA), epitomized by the IEEE 802.11 Distributed Coordination Function (DCF) \cite{802.11},\footnote{Throughout this paper we focus on plain CSMA policies,  while an investigation including collision avoidance is left as part of future work.} a terminal that has data to send picks a backoff interval whose duration $n$, in slots, is uniformly drawn in $[0,2^{\textrm{CW}-1}]$, where $\textrm{CW} = \textrm{CW}_{start} + i$, CW$_{start}$ is a system parameter to handle congestion, $i = 0, \ldots, \textrm{SRL}-1$, and SRL is the Short Retry Limit, i.e., the maximum number of transmission attempts performed at the MAC layer before dropping a packet. During this interval, the node senses the aggregate power level on the channel. If the value exceeds a given threshold $\Lambda$, the terminal stops the countdown and freezes the backoff until the medium is sensed idle again for at least a Distributed coordination function InterFrame Space (DIFS) period. On the other hand, once the backoff expires the node transmits its packet. If the destination succesfully decodes the payload, an acknowledgement (ACK) is sent to the source and the communication comes to an end.  Conversely, if the reception fails, no feedback is sent; the source increases its CW counter, and further attempts are performed after newly drawn random backoffs until either the packet is successful or the SRL is reached. 

The simple sensing mechanism implemented by CSMA allows nodes to acquire information on their neighbors' activity, and is intended both to protect ongoing communications from interference, and to block terminals whose transmissions would  fail with high probability. The effectiveness of this smart and distributed approach, however, is hindered by channel impairments such as noise, fading and shadowing as well as by the well known hidden and exposed terminal problems~\cite{Tobagi75}. Nevertheless, in light of its diffusion, it represents a significant and insightful test environment for cooperative protocols.

In order to keep our scenario general and to have a framework compatible with many other theoretical studies, we model wireless links considering Rayleigh fading, so that the received power $\eta_{n_1,n_2}$ over the channel between two nodes $\bm N_1$ and $\bm N_2$ is an exponential random variable with mean $P \delta_{n_1,n_2}^{-\alpha}$, where $P$ is the transmission power, $\delta_{n_1,n_2}$ is the Euclidean distance between the two terminals and $\alpha$ is the path loss exponent. Moreover, we consider a channel capacity model for packet decoding, so that the failure probability in retrieving an $L$-bit payload for a transmission that starts at $t_s$ and lasts for $T$ seconds can be expressed as Pr$\left\{ \mathcal L < L\right\}$, where $\mathcal L$ is the number of decoded information bits given by:
\begin{equation}
 \mathcal L = \int_{t_s}^{t_s+T} \mathcal C(\gamma(t)) dt \,,
\label{eq:decodedInfoBits}
\end{equation}
and $C(\gamma(t)) = B \log_2(1 + \gamma(t))$ is the instantaneous link capacity between sender and receiver with bandwidth $B$ and Signal to Interference and Noise Ratio (SINR) $\gamma(t)$.

Throughout this paper, we denote scalars with regular font, whereas vectors are represented in bold. Moreover, the notation $\mathcal K^{x}$ indicates quantity $\mathcal K$ conditioned on the variable $x$, and with $\mathcal B^c$ we indicate the complementary of a set $\mathcal{B}$. 

%% file: preAllocDescription.tex
\section{Proactive Cooperative Protocols in CSMA Based Ad Hoc Networks}
\label{sec:preAllocDescription}

As discussed in Section~\ref{sec:introduction}, this paper investigates the interactions between proactive cooperation and link layers based on carrier sense multiple access. With reference to a topology composed by a source $\bm S$, a destination $\bm D$ and a cooperating terminal $\bm C$, assume that some knowledge on the SINRs that would characterize data exchanges over the $\bm S$-$\bm D$, $\bm S$-$\bm C$ and $\bm C$-$\bm D$ channels is available at the nodes. In these conditions, according to the \emph{proactive cooperative} paradigm, $\bm S$ decides whether to transfer its payload directly to the intended addressee at the highest sustainable rate or to split the communication into two parts with the aim of minimizing the overall transmission time. In the latter case, during the first phase the information is delivered to the cooperator, while the destination caches the sent packet even if the used bitrate is too high for it to successfully retrieve the data content. The relay, in turn, relies on rate adaptation during the second phase to send some form of redundancy to the final addressee so as to enable decoding of the original payload. Such an approach, on the one hand, implements the key principles of cooperative relaying, with terminal $\bm C$ sending additional information to $\bm D$ on behalf of the original source of the packet to the aim of improving the quality of the wireless link. On the other hand, the described solution significanlty differs from the \emph{reactive} cooperative schemes that have been thoroughly studied in \cite{part1}, as it pre-emptively determines the opportunity to offer relayed suppot for a data exchange, rather than triggering it in the event of a communication failure over the $\bm S$-$\bm D$ channel.

We start our study by presenting two simple protocols, referred to as CSMA-CSI and Coop-CSI, that, following the principles of carrier sense medium access control described in Section~\ref{sec:csma}, take advantage of Channel State Information (CSI) to extend plain CSMA and to implement the discussed relaying approach, respectively. From this viewpoint, a first and key remark is that the effectiveness of a link layer implementing proactive cooperation is tightly correlated to the accuracy of the channel knowledge available at the nodes. It is clear that in a realistic scenario, and in particular with a fully decentralized framework, the retrieval of up-to-date information on channel qualities is not feasible due to the large amount of overhead that would be required, especially when the source has to choose among several relay candidates. On the other hand, a practical and interesting solution that tries to overcome these issues has been provided in \cite{Panwar07}, where a carrier sense-based infrastructured wireless ad hoc network is considered. The authors investigate a scenario where all nodes contend for a shared channel to deliver data to a common access point. In the proposed protocol, each source has a predefined cooperator, that is identified exploiting channel knowledge gathered from past transmissions rather than triggering dedicated negotiation procedures. This strategy has been shown to enable remarkable gains in centralized environments, yet its efficiency may be severely hindered in the non-infrastructured and non-hierarchical networks with all-to-all traffic that we consider. In fact, in our scenarios each terminal has multiple possible destinations, and thus it selects a particular addressee less frequently than in the centralized case. This implies that a knowledge of the achievable rate associated to a link learnt from previous transmissions is more likely to be outdated. Moreover, the presence of multiple simultaneous communications within the network area may induce rapidly varying interference conditions, further hampering solutions that rely on a fixed cooperator.

Starting from these remarks, in our study we assume that all the up-to-date channel knowledge needed for setting up a cooperative link is always available at nodes without any overhead. The merit of this approach is twofold. On the one hand, the solutions that will be presented and discussed represent a bound for the class of protocols that implement proactive cooperation in non-centralized ad hoc networks, and thus they allow to draw broadly applicable conclusions. On the other hand, neglecting the drawbacks that would stem at the link layer from procedures implemented to gather CSI better emphasizes the impact of issues related to carrier sense-based medium access on relaying strategies.

\subsection{Carrier Sense Multiple Access with CSI: CSMA-CSI}
\label{sec:CSMA-CSI}
Let us consider a source $\bm S$ that has data to deliver to a destination $\bm D$. As in the plain IEEE 802.11 DCF, with CSMA-CSI the node performs the backoff mechanism described in Section~\ref{sec:csma}. Once channel access has been granted, say at time $t_0$, $\bm S$ evaluates the maximum information bitrate $\rho_{s,d}$ it can use for reliably communicating with $\bm D$ based on its knowledge of the instantaneous SINR $\gamma_{s,d}(t_0)$. In order to cope with potential fluctuations that may affect the SINR over the transmission of the $L$-bit payload, e.g., due to changes in the aggregate interference perceived at the destination or due to the variability of the channel coefficient, the source follows a conservative approach, determining the sustainable bitrate as if $L(1 + \varepsilon)$ information bits had to be retrieved at $\bm D$. Recalling (\ref{eq:decodedInfoBits}), $\rho_{s,d}$ is computed as:
\begin{equation}
\rho_{s,d} = \frac{\mathcal C(\gamma_{s,d}(t_0))}{(1 + \varepsilon)} \;,
\label{eq:directLink}
\end{equation}
for a transmission time of $T_{s,d} = L/\rho_{s,d}$ seconds.

Once this calculation has been performed, the source checks whether the $\bm S$-$\bm D$ channel satisfies a minimum quality constraint, so as to avoid communications over links characterized by extremely poor conditions that would result in an inefficient usage of the bandwidth. From this viewpoint, if $\rho_{s,d} < \rho_{min}$, or, equivalently, if $T_{s,d} > T_{max}$, the node refrains from transmitting and behaves as if the communication had failed, i.e., it increases the counter of the attempts performed for the current packet and reinitiates the backoff procedure. Conversely, if $\rho_{s,d} \geq \rho_{min}$ ($T_{s,d} \leq T_{max}$), $\bm S$ transmits the payload at the maximum sustainable information bitrate and waits for a reply from $\bm D$. In the event of a successful reception, the destination replies with an ACK sent at rate $\rho_{ctrl}$ and the communication comes to an end. Otherwise, no feedback is provided, and the source iterates the described mechanism until either the payload is delivered or the SRL is reached.

\subsection{Cooperative CSMA with CSI: Coop-CSI}
\label{sec:Coop-CSI}

With reference to the transmission of an $L$-bit data unit over the $\bm S$-$\bm D$ link, let $\mathcal R_{s} = \{ \bm C_{1}, \bm C_{2},\;\dots,\; \bm C_{n} \}$ be the set of neighbors of the source node.\footnote{By neighbors we generally mean terminals that may support a node in the cooperative process. As an example, the set $\mathcal R_{s}$ may be determined at $\bm S$ by keeping track, possibly in a dynamic fashion, of the nodes it can reliably communicate with.}
As in the non-cooperative case, $\bm S$ initiates its transmission process by performing channel sensing and by computing the minimum sustainable transmission time $T_{s,d}$ over the direct link with the destination. However, when Coop-CSI is used as medium access policy, the source also checks whether it is possible to reliably deliver the payload in a shorter time by relying on the cooperating help of a surrounding terminal, splitting the transmission in two successive phases. Following this approach, the first part of the communication is devoted to a reliable data transfer from $\bm S$ to one of its neighbors. The destination, in turn, caches such a packet even if the used bitrate was too high for it to successfully retrieve the information content. Conversely, the second phase of the transmission is reserved to the relay node, which sends a different part of the original codeword obtained re-encoding the data unit sent by $\bm S$, and providing the redundancy needed for payload decoding at $\bm D$ according to a distributed Incremental Redundancy Hybrid ARQ rationale (IR HARQ), \cite{part1,harq1,harq2}.

From this perspective, a node $\bm C_{i} \in \mathcal R_{s}$ is considered as a candidate for relaying if two conditions are met at the time at which the source wants to access the channel: i) $\bm C_{i}$ is available for communication, i.e., it is not involved in any other ongoing link, and ii) $\bm C_{i}$ senses the medium idle, i.e., it is allowed to transmit. Let now $\mathcal R'_{s} \subseteq \mathcal R_{s}$ be the set of terminals that satisfy these requirements. In order to detect whether a two-hop link\footnote{We remark that, throughout this paper, the term \emph{two-hop} link refers to the proactive cooperative paradigm, with the destination caching the packet sent by $\bm S$ during the first phase and with the relay sending redundancy on $\bm S$'s packet during the second phase, and is not to be intended in the classical routing sense.} that offers improvement over the direct transmission is available, the source starts by computing for each $\bm C_{i} \in \mathcal R_{s}'$ the maximum sustainable information bitrate for the first phase of a cooperative communication, based on the current value of the SINR $\gamma_{s,c_{i}}(t_0)$:
\begin{equation}
\rho_{s,c_i} = \frac{\mathcal C(\gamma_{s,c_{i}}(t_0))}{1 + \varepsilon} \;.
\label{eq:directLink_sc}
\end{equation}
The corresponding transmission time is $T_{s,c_{i}} = L/\rho_{s,c_{i}}$ and, according to (\ref{eq:decodedInfoBits}), the destination may exploit the $\bm S$-$\bm C_{i}$ data exchange to retrieve $\mathcal L_{1,i}= T_{s,c_{i}} \, \mathcal C(\gamma_{s,d}(t_0))$ information bits.\footnote{Note that the computation of $L_{1,i}$ implicitly assumes the channel capacity between $\bm S$ and $\bm D$ to remain constant for the whole transmission. This simplification is necessary, as the source has to choose which policy to trigger based only on CSI at time $t_0$. In this perspective, non-optimal decisions that may stem from fluctuations of $C(\gamma_{s,d}(t))$ for $t>t_0$ are unavoidable due to the causality of the system.} Therefore, the overall duration of the two-hop solution with $\bm C_{i}$ as cooperator can be computed as:
\begin{equation}
T_{split,i} = T_{s,c_{i}} + \frac{L - \mathcal
L_{1,i}}{\rho_{c_{i},d}} \;, \label{eq:tSplit}
\end{equation}
where the second term represents the time needed for the $\bm C_{i}$-$\bm D$ transmission to allow reliable decoding at the destination, and where  $\rho_{c_{i},d} = \mathcal C(\gamma_{c_{i},d}(t_0)) / (1 + \varepsilon)$ is an estimate of the maximum sustainable bitrate during the second phase of the communication.

Upon completing these calculations, the source decides in which way to deliver its payload evaluating
\begin{equation}
T^{*} = \displaystyle \min_{\bm C_{i} \in \mathcal R_{s}'} \left \{ T_{split,i}, T_{s,d}, T_{max} \right\}\;.
\label{eq:minT}
\end{equation}
In particular, if $T^{*} = T_{max}$, no solution providing sufficient reliability is available. In this case, the node refrains from transmitting, increases the counter for the number of attempts performed for the current packet and reinitiates the backoff procedure. Conversely, if $T^{*} = T_{s,d}$, $\bm S$ simply sends the data unit over the direct link at rate $\rho_{s,d}$, and the communication follows the principles of CSMA-CSI. Finally, when $T^{*} = T_{split,j}$, with $j = \textrm{argmin}_i \{ T_{split,i} \}$, the source starts the data exchange by sending the payload to $\bm C_{j}$ at rate $\rho_{s,c_{j}}$ computed through (\ref{eq:directLink_sc}). Notice that the header of such packet contains information describing the structure of the transmission, i.e., source, destination, chosen relay and ID of the data unit. In this way, if the final addressee decodes the header, a corrupted version of the incoming data can properly be cached, and the terminal becomes aware that additional redundancy will be sent over a cooperative link.\footnote{Due to sudden changes in the capacity of the $\bm S$-$\bm D$ link, the destination may decode the payload even if it was sent at rate $\rho_{s,c_j}$. However, this case, though possible, is not likely to happen, and is disregarded in our studies.} Therefore, as soon as $\bm C_{j}$ retrieves the information sent by $\bm S$, the payload is re-encoded so as to form a different part of the original codeword, and the relay immediately delivers it to $\bm D$ with a transmission at rate $\rho_{c_{j},d}$. If the destination decodes the data unit, an acknowledgement is sent, and the communication comes to an end. Otherwise, the source iterates the described procedure until either successful delivery is achieved or the SRL is reached.

In conclusion, let us make some remarks on the presented cooperative protocol. In the first place, we shall notice that the proactive paradigm could be implemented also with techniques different from IR HARQ, e.g., with the relay transmitting a whole copy of the packet of $\bm S$  and with $\bm D$ performing maximum ratio combining on two instances of the same frame. Nonetheless, the discussion carried out in \cite{part1} highlights that IR HARQ appears as the optimal scheme to cope with some specific issues that arise in carrier sense-based environments, and thus represents an ideal solution for our investigations.
Secondly, we remark that besides relying on perfect channel and interference knowledge as its non-collaborative counterpart, Coop-CSI also idealizes some aspects of medium access contention, as we assume all the terminals involved in a data exchange to be immediately and seamlessly aware of how the communication will be structured. On the one hand, such an assumption falls under the line of reasoning, discussed at the beginning of this section, of dealing with schemes that represent a bound for the class of protocols that implement proactive cooperation, so as to draw conclusions that are both broadly applicable and stem from the intrinsic interactions between carrier sensing and cooperation rather than by specific implementation details. On the other hand, we also notice that the proposed protocol does not embody the proactive relaying paradigm at too abstract a  level, as Coop-CSI could easily be modified in order to be practically implemented, e.g., by following the approaches presented in \cite{Panwar07,Shan09}.

%% file: preAllocAnalysis.tex
\section{An Analytical Framework for the Study of Proactive Cooperation in CSMA Networks}
\label{sec:preAllocAnalysis}

We start our analysis of schemes that implement proactive
cooperation by investigating the performance gain they can offer
over basic solutions when nodes have to obey no medium access
control policy, i.e., assuming perfect coordination among terminals.
To this aim, we focus on the protocols described in
Section~\ref{sec:preAllocDescription} disregarding the constraints
of CSMA. In particular, we consider for the moment a topology
composed by a source $\bm S$ and a destination $\bm D$ placed in a
region $\mathcal A$ at positions $p_s = \{x_s, y_s \}$ and $p_d =
\{x_d, y_d\}$ respectively, and by a relay terminal $\bm C$ deployed
in the middle of the line connecting them.\footnote{As will be
discussed later, this configuration maximizes the gains offered by
proactive cooperation.} For the sake of mathematical tractability,
following the traditional approach used in theoretical works, e.g.,
\cite{Laneman-dec04,Lai06}, we assume fading coefficients to remain
constant for the whole duration of a communication\footnote{The
assumption of constant fading coefficients will be relaxed in the
simulation studies presented in this work by accurately modeling the
time-correlated nature of the wireless channel, see
Section~\ref{sec:preAllocSims}.} and we do not model a source of
interference subject to carrier sense, but we rather assume an
interference level with average power $\sigma^2_i$ at both $\bm D$
and $\bm C$.

The basic transmission policy that we analyze, as discussed, always
uses the direct link to deliver data from the source to the
destination. The maximum sustainable rate in this case is given by
$\rho_{s,d} = \mathcal C(\gamma_{s,d})$, with $\gamma_{s,d} =
\eta_{s,d}/(N + \iota_d)$.\footnote{The margin factor $\varepsilon$
discussed in Section~\ref{sec:preAllocDescription} is not considered
here, since fading and interference are assumed to remain constant
for the whole duration of a transmission.} Here, both the desired
received power $\eta_{s,d}$ and the interfering power $\iota_d$ at
the destination are modeled as exponential random variables, with
mean values $P \delta_{s,d}^{-\alpha}$ and $\sigma^2_i$,
respectively. The throughput $\tau_{direct}$ offered by this
approach can be computed by averaging the achievable information
bitrate over fading and interference, obtaining:
\begin{equation}
 \tau_{direct} \!=\! \frac{P \delta_{s,d}^{-\alpha}}{P \delta_{s,d}^{-\alpha} - \sigma_i^2} \, \left[ \mathcal G\left( -\frac{N}{P \delta_{s,d}^{-\alpha}}\right) - \mathcal G \left(
-\frac{N}{\sigma^{2}_{i}} \right)\right] \;,
\label{eq:truDirect}
\end{equation}
where $N$ is the noise floor, and $\mathcal G(a)$ is defined as:
\begin{equation}
 \mathcal G(a) = \frac{B}{\ln(2)} \, e^{-a} \, \int_{-a}^{+\infty} \frac{e^{-t}}{t} dt \;.
\end{equation}

On the other hand, the cooperative protocol chooses whether to employ a direct or a two-hop link so as to minimize the overall transmission time. In this case, the sustainable bitrates involving the relay terminal are given by $\rho_{s,c}= \mathcal C(\gamma_{s,c})$ and $\rho_{c,d}= \mathcal C(\gamma_{c,d})$, and the channel capacities are determined by the SINRs $\gamma_{s,c} = \eta_{s,c}/(N + \iota_c)$ and $\gamma_{c,d} = \eta_{c,d}/(N + \iota_d)$, where $\eta_{s,c}$, $\eta_{c,d}$ and $\iota_c$ are again modeled as exponential variables with mean values $P \delta_{s,c}^{-\alpha}$, $P \delta_{c,d}^{-\alpha}$ and $\sigma^2_i$, respectively. Applying (\ref{eq:tSplit}), the time required for data delivery over a cooperative link can easily be computed as:
\begin{equation}
  T_{split} = L \, \frac{\mathcal C(\gamma_{s,c}) + \mathcal C(\gamma_{c,d}) - \mathcal C(\gamma_{s,d})}{\mathcal C(\gamma_{s,c}) \, \mathcal C(\gamma_{c,d})}\;.
\label{eq:tSplitAnalysis}
\end{equation}
Starting from this result, and introducing the random vector
$\mathbf v = \{\eta_{s,d}, \eta_{s,c}, \eta_{c,d}, \iota_c,
\iota_d\}$,two regions can be identified: $\Delta_{split}(\mathbf v)
= \left\{ \mathbf v \; | \; \mathcal C(\gamma_{s,c}) \geq \mathcal
C(\gamma_{s,d}), \mathcal C(\gamma_{c,d}) \geq \mathcal
C(\gamma_{s,d}) \right\}$ and $\Delta_{direct}(\mathbf v) =
\Delta_{split}^c(\mathbf v)$, so that for $\mathbf v \in
\Delta_{split}$ a relayed communication is performed, whereas for
$\mathbf v \in \Delta_{direct}$ the protocol resorts to the direct
link. It follows that the average throughput $\tau_{coop}$ for the
cooperative solution can be computed as:
\begin{equation}
 \tau_{coop} = \int_{\Delta_{split}} \frac{\mathcal C(\gamma_{s,c}) \, \mathcal C(\gamma_{c,d})}{\mathcal C(\gamma_{s,c}) +
 \mathcal C(\gamma_{c,d}) - \mathcal C(\gamma_{s,d})} \; f_{\mathbf v}(\mathbf v) d\mathbf v
+ \int_{\Delta_{direct}} \mathcal C(\gamma_{s,d}) \; f_{\mathbf v}(\mathbf v) d\mathbf v  \,,
\label{eq:truCoop}
\end{equation}
where $f_{\mathbf v}$ is the joint probability density function of vector $\mathbf v$, factorizable as the product of the densities of the involved exponential variables,\footnote{Such a decomposition holds since all the components of $\bm v$ involve spatially disjoint, and therefore statistically independent, channels.} and where the integrand over $\Delta_{split}$ has been obtained from (\ref{eq:tSplitAnalysis}).

\begin{figure}
\centering
\includegraphics[width=\figw]{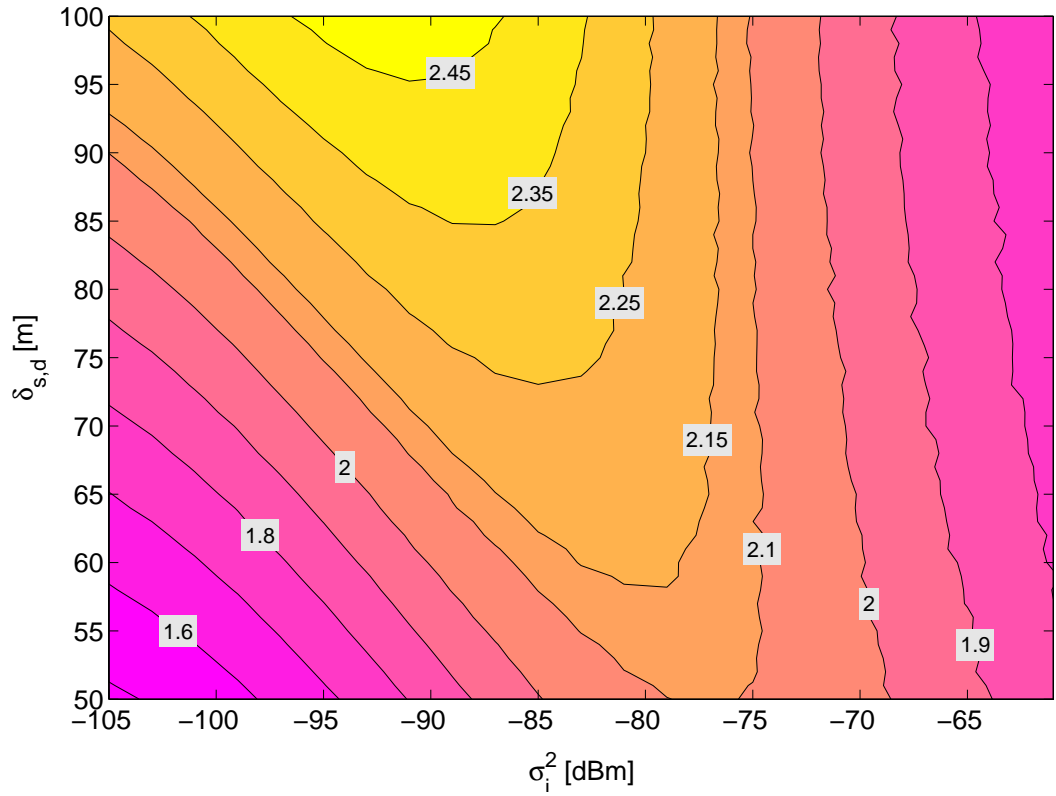}
\caption{Average throughput gain of cooperative over non-cooperative solutions for different values of $\delta_{s,d}$ and for different interference levels when no medium contention and uniform interference are considered. The cooperator is assumed to be located in the middle of the line connecting $\bm S$ and $\bm D$.}
\label{fig:preAnalysis_molare}
\end{figure}

\begin{table}[b]
  \begin{center}
  \caption{Parameters used in our studies}
  \label{tab:parameters}
  \begin{tabular}{|c|c|}
  \hline
  Transmission power (dBm) & 10 \\
  \hline
  Noise Floor (dBm) & -102 \\
  \hline
  Detection threshold (dBm) & -96\\
  \hline
  Path loss exponent, $\alpha$ & 3.5\\
  \hline
  Maximum Doppler frequency, $f_d$ (Hz) & 11.1\\
  \hline
  Carrier Frequency (GHz) & 2.4\\
  \hline
  Bandwidth, $B$ (MHz) & 1\\
  \hline
  Headers and control pkt information bitrate, $\rho_{ctrl}$ (Mbps) & 0.532\\
  \hline
  Minimum src-dest capacity for proactive coop, $\rho_{min}$ (Mbps)& 0.95\\
  \hline
  CS threshold, $\Lambda$ (dBm) & -100\\
  \hline
  Slot, DIFS, SIFS duration ( $\mu$s)& 10, 128, 10\\
  \hline
  Initial contention window index, CW$_{start}$ & 5 \\
  \hline
  SRL - Coop-CSI, CSMA-CSI & 4, 5 \\
  \hline
  Interference margin for proactive cooperation, $\varepsilon$ & 0.15 \\
  \hline
  DATA header (bit) & 112 \\
  \hline
  Payload (bit) & 5000 \\
  \hline
  ACK/NACK (bit) & 112 \\
  \hline
  \end{tabular}
  \end{center}
\end{table}

Fig.~\ref{fig:preAnalysis_molare} reports the ratio of $\tau_{coop}$ to $\tau_{direct}$ with network and protocol parameters set as in Tab.~\ref{tab:parameters}. Here, the $x$ axis represents the average interference perceived at $\bm C$ and $\bm D$, i.e., $\sigma^2_i$, while the $y$ axis indicates the distance $\delta_{s,d}$ between source and destination. As a general remark, the plot clearly shows that proactive cooperation has the potential to enable important improvements over simpler communication schemes in small topologies and when medium access is perfectly coordinated among nodes. Under these assumptions, indeed, the average throughput offered by relaying strategies is more than twice that achieved by basic solutions for a broad range of networking parameters. Moreover, it is possible to observe how, for a given distance between source and destination, relaying is exploited at its utmost for intermediate values of $\sigma^2_i$. The reason is that for very low interference the direct link already offers extremely good performance and thus there is little room for improvement. On the other hand, when $\sigma_i^2$ raises above a critical value not only does the capacity offered by the direct link deteriorate, but also the quality of the $\bm S$-$\bm C$ and $\bm C$-$\bm D$ channels plummets, dilating the time required for a two-hop communication more than proportionally and consequently shrinking the achievable throughput gain.
Finally, Fig.~\ref{fig:preAnalysis_molare} also suggests that, when a fixed level of interference is considered, cooperation performs better for larger values of $\delta_{s,d}$, as higher path-losses affecting the direct link favor data transfers over two faster hops.

We now broaden our investigation by focusing on how carrier
sense-based medium access can impact proactive relaying strategies.
To this aim, we refer again to a scenario composed by a source $\bm
S$, a destination $\bm D$, an interferer $\bm I$ and a potential
cooperator $\bm C$ placed within a region $\mathcal A$ at positions
$p_s = \{x_s,y_s\}$, $p_d = \{x_d,y_d\}$, $p_i = \{x_i,y_i\}$ and
$p_c = \{x_c,y_c\}$, respectively. Furthermore, for the sake of
mathematical tractability, we assume that $\bm I$ accesses the
channel simultaneously with $\bm S$ if not locked by carrier sense,
and that its transmission lasts for the whole duration of the
communication between the source and the destination.

A first important observation regards the bias induced by CSMA on the position of the interfering node. Recalling the system model presented in Section~\ref{sec:csma}, the probability $\mathcal F(p_s,p_i)$ that a terminal located at $p_{i}$ is allowed to access the channel given a transmission being performed by $\bm S$ is given by
\begin{equation}
\mathcal{F}(p_s,p_i) = \textrm{Pr}\{ \eta_{s,i} + N < \Lambda \} = 1 - e^{-\frac{\Lambda - N}{P \delta_{s,i}^{-\alpha}}} \,.
\label{eq:csConstraint}
\end{equation}

$\mathcal F(p_s,p_i)$ exhibits a radial symmetry centered in $p_s$,
with an exponentially decreasingprobability of sensing the medium as
idle for larger values of $\delta_{s,i}$. It follows that, as
expected, $\bm I$ tends to be concentrated far away from the source.
In the perspective of our discussion it is important to notice that,
while this distribution is meant to protect the direct link so as to
enhance its decoding probability, it also severely hinders the
effectiveness of cooperative mechanisms. In order to understand this
effect, let us recall that a relayed transmission with the help of
$\bm C$ is triggered according to Coop-CSI only if two conditions
are met: i) $\bm C$ senses the medium idle, and ii) the time
required for delivering data over the $\bm S$-$\bm C$-$\bm D$ path
is shorter than the duration achievable by means of a direct link
between source and destination, i.e., $T_{split} < T_{s,d}$.

Consider first the former requirement. Conditioned on the position of the interferer, the probability that a node in $p_c$ is allowed to access the medium can be expressed as $\mathcal F(p_i, p_c)$. Therefore, letting $\bm p = \{p_s,p_d,p_c \}$ be the topology vector under consideration, the probability $\mathcal M(\bm p)$ that $\bm C$ meets requirement i) can be obtained by averaging the conditional value over the spatial distribution of $\bm I$, obtaining:
\begin{equation}
 \mathcal M(\bm p) = \frac{\int_{\mathcal A} \mathcal F(p_i, p_c) \; \mathcal F(p_s, p_i) \;dp_i}
{\int_{\mathcal A} \mathcal F (p_s, p_i) \; dp_i}
\label{eq:pFreeCoop}
\end{equation}

\begin{figure}
\centering
\includegraphics[width=\figw]{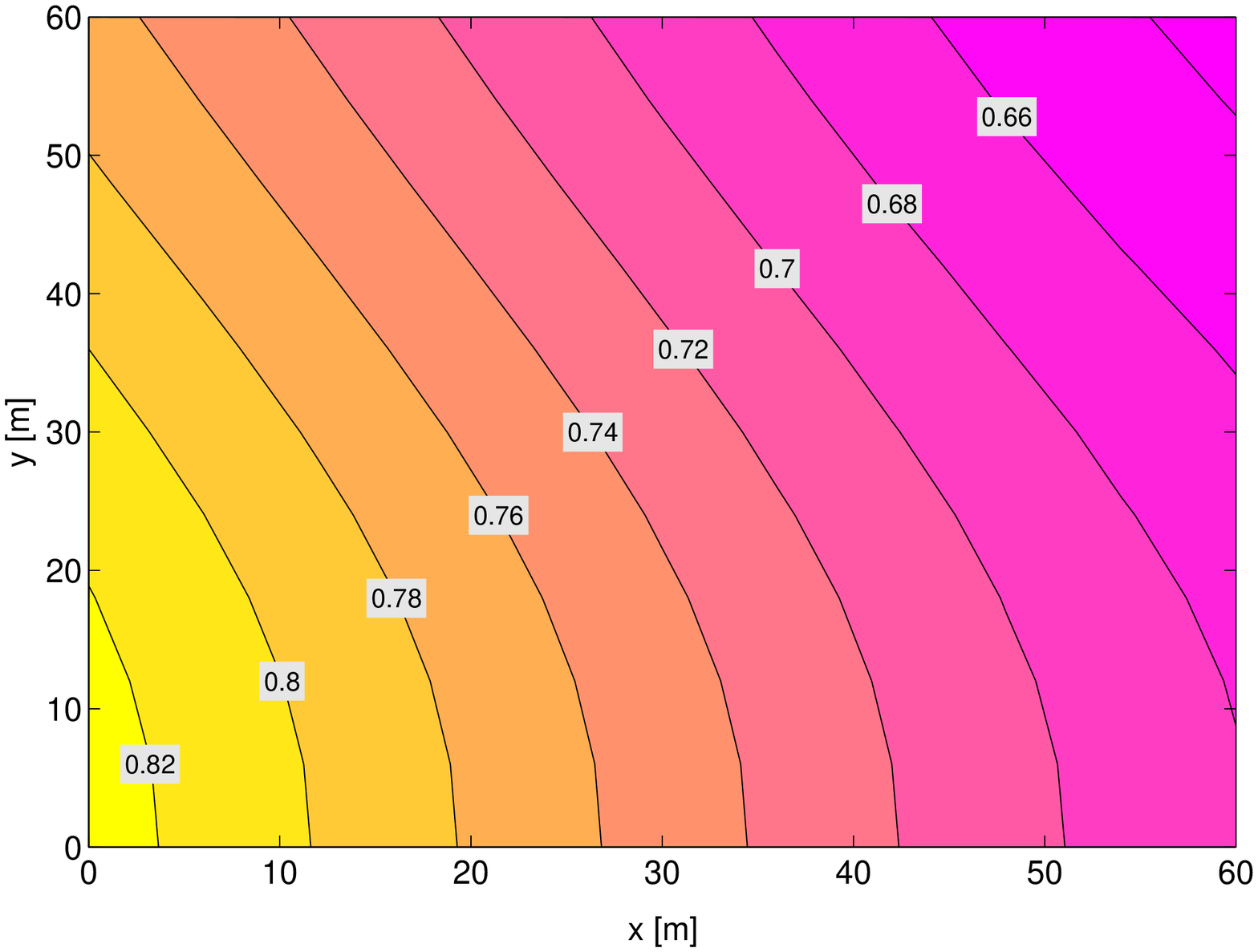}
\caption{Probability that a node in $p_{c} = \{x,y\}$ senses the medium idle in a proactive cooperative scenario. $p_{s} = \{0,0\}$, $p_{d} = \{60,0\}$.}
\label{fig:preAnalysis_pFreeCoop}
\end{figure}

The results of the numerical evaluation of (\ref{eq:pFreeCoop}) for a region $\mathcal A$ spanning $x\in[-150, 200]$, $y\in[-200,200]$, a topology $p_s = \{0,0\}$, $p_d = \{60, 0 \}$ (all the coordinates being expressed in meters), and parameters set according to Tab.~\ref{tab:parameters} are shown in Fig.~\ref{fig:preAnalysis_pFreeCoop}.\footnote{While $\mathcal A$ has been chosen wide enough to also take into account the influence of interfering nodes located far from the source-destination pair, the figures shown here plot the identified distributions restricted to only a part of the region for the sake of clarity.} The plot clearly highlights how terminals allowed to take part in a relaying phase while obeying the principles of CSMA are concentrated in the proximity of the source, as a consequence of the lower interference level induced by the medium access policy in such a region.

On the other hand, focusing on requirement ii), let $\mathcal R(\bm p)$ be the probability that a cooperative communication involving $\bm C$ offers improvement over the direct link. Recalling the approach discussed in the first part of this section for the throughput derivation, $\mathcal R^{p_i}(\bm p)$, i.e., $\mathcal R(\bm p)$ conditioned on the position of the interferer, can be evaluated by simply integrating $f_{\mathbf v}(\mathbf v)$ over $\Delta_{split}(\mathbf v)$. In this case, however, the random variables $\iota_c$ and $\iota_d$, i.e., the components of $\mathbf v$ that describe the interference level perceived at $\bm C$ and $\bm D$, are no longer i.i.d. with mean value $\sigma_i^2$. Instead, we model them as exponential random variables with mean values $P \delta_{i,c}^{-\alpha}$ and $P \delta_{i,d}^{-\alpha}$, respectively, so as to capture the effects of both Rayleigh fading and the interference correlation induced by having $\bm I$ at a given position subject to the carrier sense constraint. Under these assumptions, the sought probability can be computed by averaging $\mathcal R^{p_i}(\mathbf p)$ over $p_i$, obtaining:
\begin{equation}
 \mathcal R(\bm p) = \frac{\int_{\mathcal A} \mathcal R^{p_i}(\mathbf p) \; \mathcal F(p_s, p_i) \;dp_i}
{\int_{\mathcal A} \mathcal F (p_s, p_i) \; dp_i} \;.
\end{equation}

\begin{figure}
\includegraphics[width=\figw]{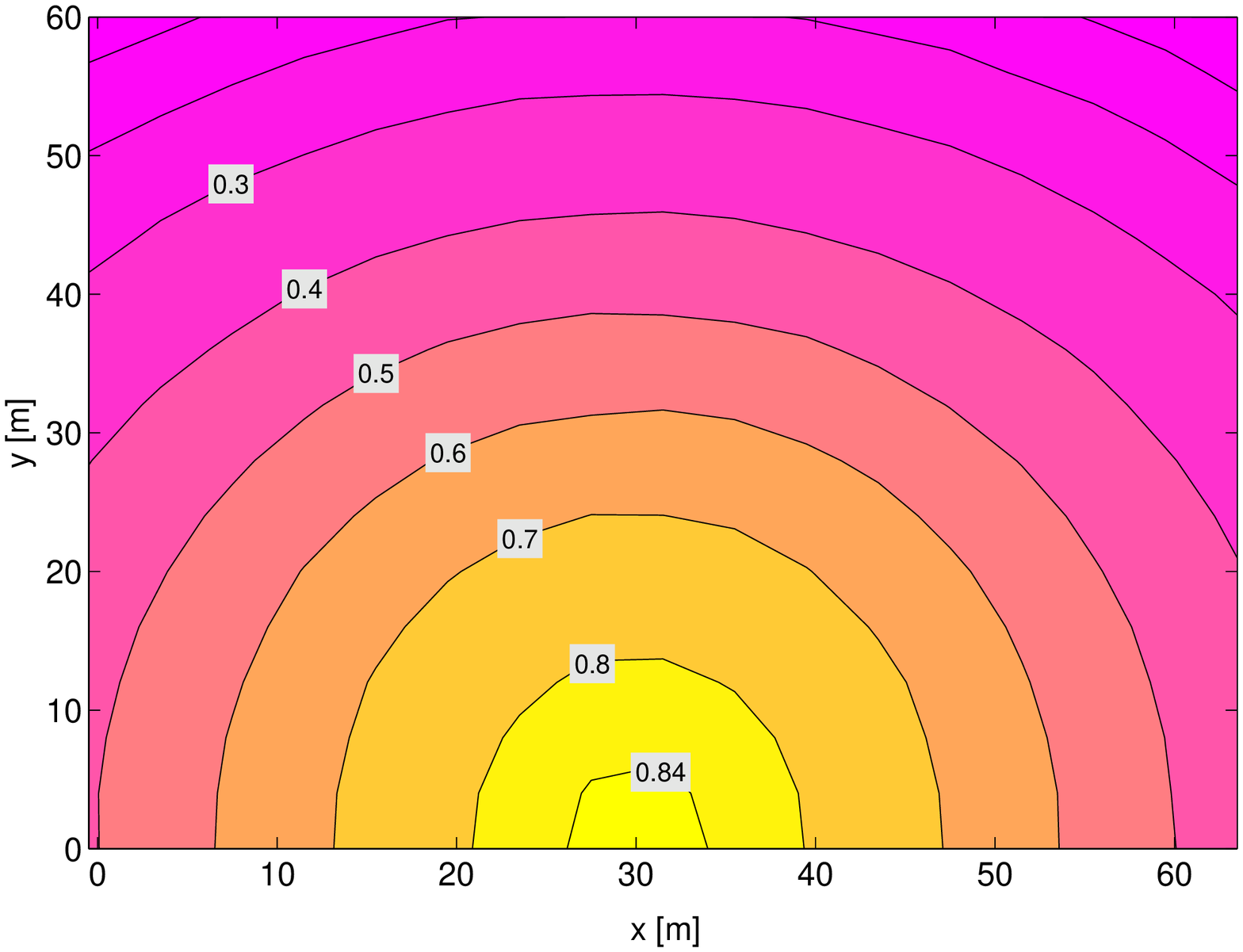}
\caption{Probability that a node in $p_{c} = \{x,y\}$ satisfies the rate constraint to trigger proactive cooperation. $p_{s} = \{0,0\}$, $p_{d} = \{60,0\}$.}
\label{fig:preAnalysis_pChosenCoop}
\end{figure}

Fig.~\ref{fig:preAnalysis_pChosenCoop} reports the values obtained for $\mathcal R(\bm p)$ in the topology under consideration. From the plot, it is apparent how cooperation is much more likely to be triggered when the relay is located in the middle of a line connecting source and destination. In such a condition, indeed, the two-hop solution benefits from both spatial diveristy and a reduced and balanced path loss for the two links. Conversely, if the cooperator is in the proximity of the source, the average powers received over the $\bm S$-$\bm D$ and the $\bm C$-$\bm D$ channels are alike, and so are the sustainable transmission times, thus limiting the achievable gains of a cooperative communication to fading diversity only. For symmetry, a similar reasoning holds when $\bm C$ is close to the final addressee of the payload.

The analytical framework presented so far makes it possible to draw some important conclusions on the interaction between CSMA and proactive cooperation. As a first observation, comparing the results of Figs.~\ref{fig:preAnalysis_pFreeCoop} and \ref{fig:preAnalysis_pChosenCoop}, we can infer that the employed medium access control policy restricts the set of relay candidates to nodes that are not likely to offer significant benefits in terms of sum rate for payload delivery. This, in turn, reduces the number of triggered cooperative phases (see Section~\ref{sec:preAllocDescription}), with a detrimental effect on the potential gains offered by relaying. Notice that such a consideration is not specific to the protocol implementation that we have proposed, but rather it intrinsically stems from the uneven interference distribution yielded by the carrier sense paradigm.

\begin{figure}
\includegraphics[width=\figw]{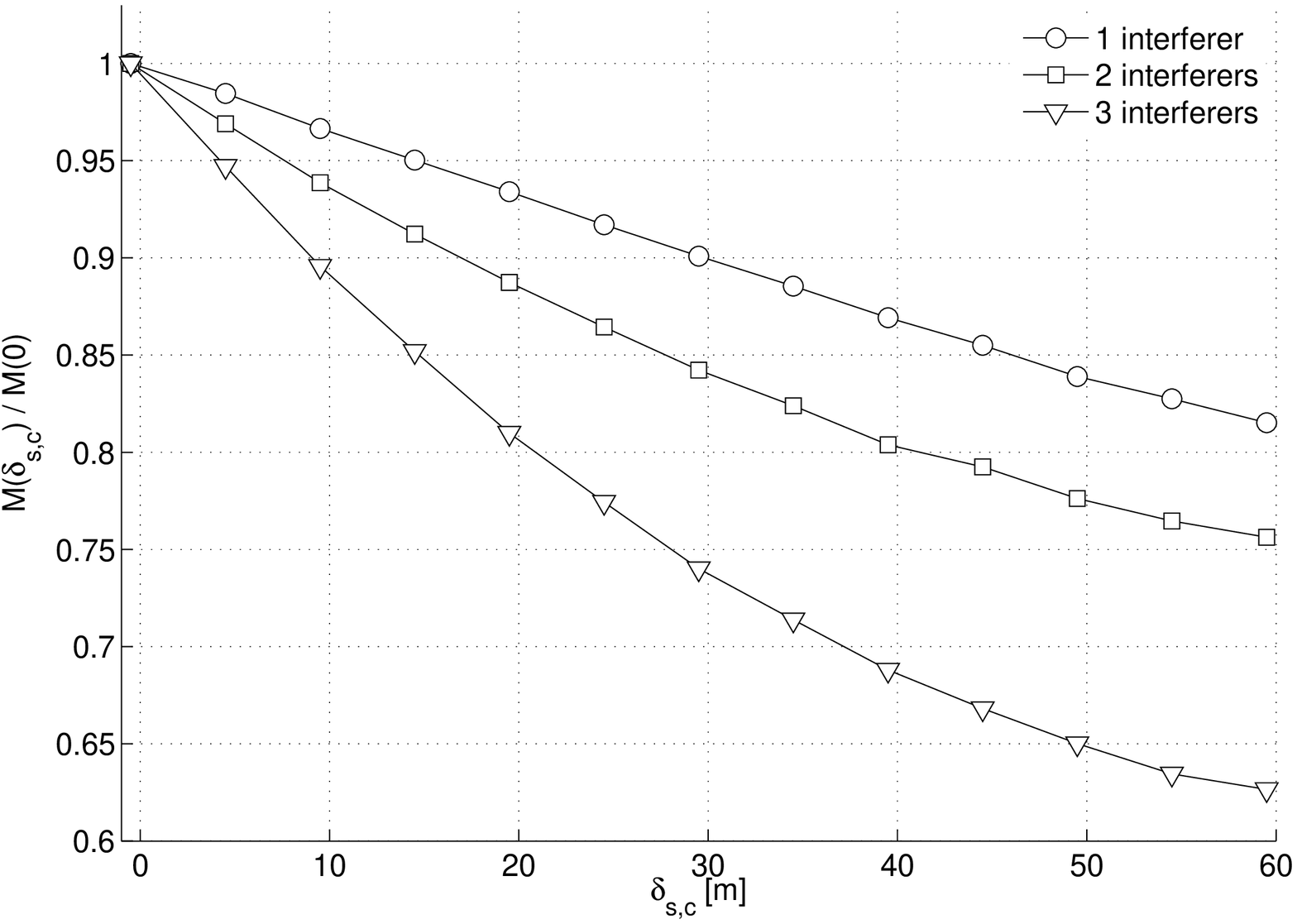}
\caption{Ratio of the probability that a node lying on the line connecting source and destination at distance $\delta_{s,c}$ from the former senses the medium idle to the same quantity computed at the position of the source. $k\in\{1,2,3\}$ interferers are distributed over region $\mathcal A$ with $\bm S$ and $\bm D$ located at $p_s = \{0,0\}$ and $p_d = \{ 60,0\}$, respectively.}
\label{fig:preAnalysis_pFreeCoopMultipleInterf}
\end{figure}

It is also interesting to remark that even though this effect is already manifest in the basic four-node topology studied up to now, its impact is exacerbated when more realistic scenarios are taken into account. We have investigated this aspect by applying the developed framework also to networks with multiple interfering terminals deployed over $\mathcal A$ and subject to CSMA. To this aim, Monte Carlo simulations have been performed, distributing interfering nodes so that each of them satisfies the carrier sense constraint both for the transmission of $\bm S$ and for the transmissions of other active interferers. For the sake of simplicity, we have considered a potential cooperator located along the line connecting $\bm S$ and $\bm D$, so that the probability $\mathcal M(\bm p)$ that $\bm C$ senses an idle medium, i.e., that it satisfies requirement i), only depends on its distance from the source. Fig.~\ref{fig:preAnalysis_pFreeCoopMultipleInterf} reports the ratio $\mathcal M(\delta_{s,c})/\mathcal M(0)$, i.e., the sought distribution normalized to the probability that a node in $p_s$ senses the medium idle despite the active interfering communications, for $p_s = \{0,0\}$ and $p_d=\{60,0\}$. The results confirm that even a limited increase in the number of interferers further concentrates relay candidates, i.e., nodes that sense the medium idle, in the proximity of the source, thus reducing the chances of enabling efficient cooperative phases. In particular, while with a single interferer (see also Fig.~\ref{fig:preAnalysis_pFreeCoop}) the probability that a relay is allowed to access the channel is just 10\% lower when located in the most favorable position for throughput maximization, i.e., $p_c=\{30,0\}$, than when positioned close to the source, the gap increases to more than 15\% and 25\% for networks with two and three interferers, respectively.

\begin{figure}
\includegraphics[width=\figw]{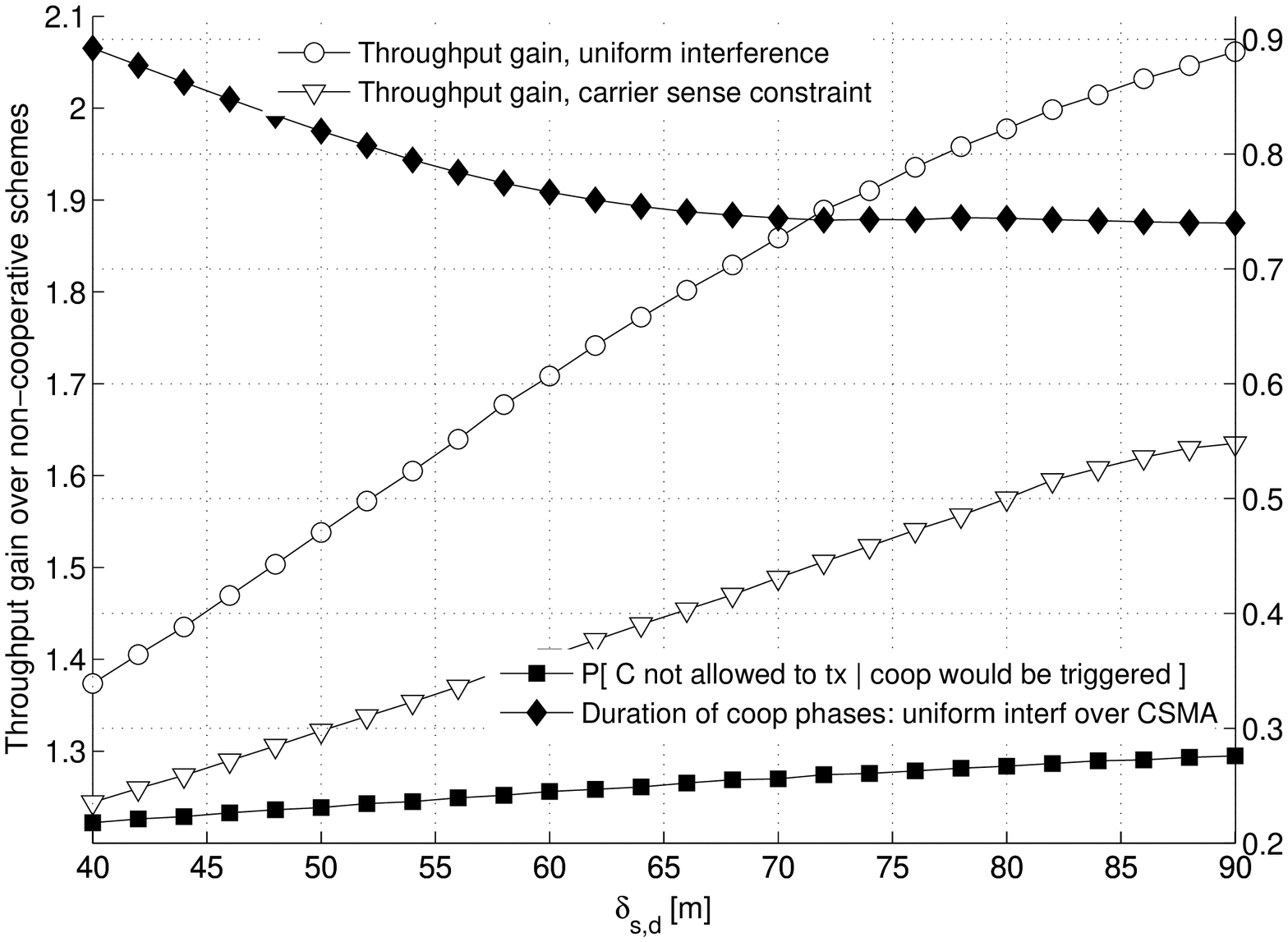}
\caption{Comparison of proactive cooperation with and without carrier sense constraints for different values of $\delta_{s,d}$. Left axis, white markers: throughput gain over non-cooperative behavior, i.e., ratio of the aggregate throughputs achieved by the two solutions. Right axis, black markers: potential cooperative phases lost because the relay senses the medium busy (square markers), and ratio of the average duration of a cooperative transmission with no MAC to the same quantity when CSMA is considered (diamond markers). }
\label{fig:preAnalysis_truputBiased}
\end{figure}

The influence of CSMA on the availability of cooperators shown by Figs.~\ref{fig:preAnalysis_pFreeCoop}, \ref{fig:preAnalysis_pChosenCoop} and \ref{fig:preAnalysis_pFreeCoopMultipleInterf} has a direct impact on the overall network performance. From this viewpoint, it is insightful to compare the throughputs achieved by the same relaying strategy under uniform and carrier sense-biased interference conditions. To this aim, let us refer again to a topology composed of a source-destination pair at distance $\delta_{s,d}$ and of a relay node uniformly distributed over region $\mathcal A$. By means of Monte Carlo simulations we have computed the average throughputs offered by the plain and cooperative strategies described in Section~\ref{sec:preAllocDescription} both when interference with average power $\sigma_{i}^{2}$ affects $\bm D$ and $\bm C$, and when an interferer $\bm I$ subject to carrier sense based medium access control is distributed over $\mathcal A$.

The curves with white markers in Fig.~\ref{fig:preAnalysis_truputBiased} report the obtained results in terms of performance gain of the relaying schemes over their non-cooperative counterparts in the two scenarios as a function of  $\delta_{s,d}$. In order to have a fair comparison, for each value of the distance between $\bm S$ and $\bm D$, $\sigma_{i}^{2}$ in the non-CSMA case has been set equal to the average interference level perceived at the destination when $\bm I$ has to obey the medium access policy. The plot clearly stresses the influence of the considered MAC, as the achievable throughput gain reduces by up to 40\% when carrier sense is implemented. Also notice that the performance gap in the two scenarios becomes more pronounced for higher values of $\delta_{s,d}$. As highlighted in the first part of our discussion (see Fig.~\ref{fig:preAnalysis_molare}), a larger distance between $\bm S$ and $\bm D$ favors relayed transmissions with respect to data delivery over direct links. Nonetheless, such a beneficial effect is counterbalanced by the biased distribution of potential relays that appears when nodes adhere to the principles of CSMA. Indeed, the farther away the destination from the source, the less the area where cooperators offer the maximum improvement is protected by the carrier sense mechanism.

It is important to observe that this performance degradation stems from two key factors: on the one hand CSMA reduces the number of cooperative phases that can take place, and on the other hand it affects their effectiveness. This inference is supported by the curves with black markers in Fig.~\ref{fig:preAnalysis_truputBiased}, whose values are to be referred to the $y$ axis on the right side of the plot. The former detrimental effect is shown by the square-marked line, that reports the fraction of times in which Coop-CSI has to refrain from relaying because a candidate that would offer a performance gain over the direct link is not allowed to access the channel. Even when $\bm S$ and $\bm D$ are close to each other, more than 20\% of the cooperative attempts have to be aborted, with such percentage increasing up to 30\% for larger values of $\delta_{s,d}$.  The diamond-marked curve, in turn, depicts the ratio of the average duration of an actually performed cooperative transmission when no medium contention and uniform interference are considered to the same quantity evaluated when CSMA is applied, showing an efficiency loss of up to almost 30\%. It is then apparent that even when cooperation takes place carrier sense hampers its potential by enforcing the usage of relays that are located at suboptimal positions.

In conclusion, the analysis carried out in this section has pointed out some fundamental repercussions induced by medium access control based on carrier sense on cooperative mechanisms when CSI is available. While these aspects are already noticeable in simple topologies, their impact may be further stressed when relaying is applied to more realistic and large networking scenarios, where additional issues related to the interaction of several nodes come into play. With a view to understanding these problems, the next section will present the results of extensive simulation campaigns that we have performed.

%% file: preAllocSims.tex
\section{Simulation Results}
\label{sec:preAllocSims}

The protocols described in Section~\ref{sec:preAllocDescription} have been tested in large and realistic networking environments by means of extensive Omnet++ \cite{Varga01} simulations. As in the companion paper \cite{part1}, several configurations have been evaluated, analyzing different system cardinalities, densities and parameters. On the other hand, due to space constraints, and in view of the similarity of the key trends that have been observed, we only report in detail the results obtained in a specific scenario. Throughout this section, thus, we refer to a network composed of 35 terminals randomly distributed in a $300 m \times 300 m$ area, with each node generating single-hop Poisson traffic addressed to its neighbors with intensity $\lambda$ (kbit/s/node). Such a configuration is particularly apt to evaluate the performance of carrier sense based link layers, as enough spatial separation among nodes is provided to support the presence of simultaneous communications in the network, leading to harsh interference, channel contention and hidden terminal conditions. The wireless environment, in line with the analytical framework developed in this paper, is subject to frequency flat Rayleigh fading. In addition, our simulations take into account the temporal coherence of the wireless medium, which is modeled according to Jakes' approach for land mobile fading \cite{Jakes93}, so that the correlation between two instances of the same channel spaced in time by $\tau$ seconds is given by $J_0(2\pi f_d \tau)$, where $J_0(\cdot)$ is the zero-order Bessel function of the first kind and $f_d$ is the maximum Doppler frequency. The information bitrate for headers and control packets has been set to 0.53 Mb/s, so as to guarantee a decoding probability of 0.95 at a distance of 60 $m$ with a single transmission, whereas the minimum channel capacity $\rho_{min}$ required to perform a transmission attempt has been set to 0.95 Mb/s. To have a fair comparison between CSMA-CSI and Coop-CSI, we have tuned the protocol parameters so that they offer a similar reliability. In particular, a Packet Delivery Ratio of 95\% has been obtained by choosing a Short Retry Limit of 5 for the plain CSMA scheme, whereas 4 attempts were enough for the cooperative solution, in light of the spatial diversity gain it leverages. In addition, the margin coefficient $\varepsilon$ (see Section~\ref{sec:preAllocDescription}), which intrinsically depends on the variability of the wireless environment and induces a critical tradeoff between transmission time and decoding probability, has been set to 0.15 after careful studies so as to maximize the aggregate throughput.
All the other protocol and network parameters used in our studies are reported in Tab.~\ref{tab:parameters}. The results presented in this paper have been obtained by averaging the outcome of 50 independent simulations, so that the 95\% confidence intervals, although not reported for readability, never exceed 3\% of the estimated value.

\begin{figure}
\centering
\includegraphics[width=\figw]{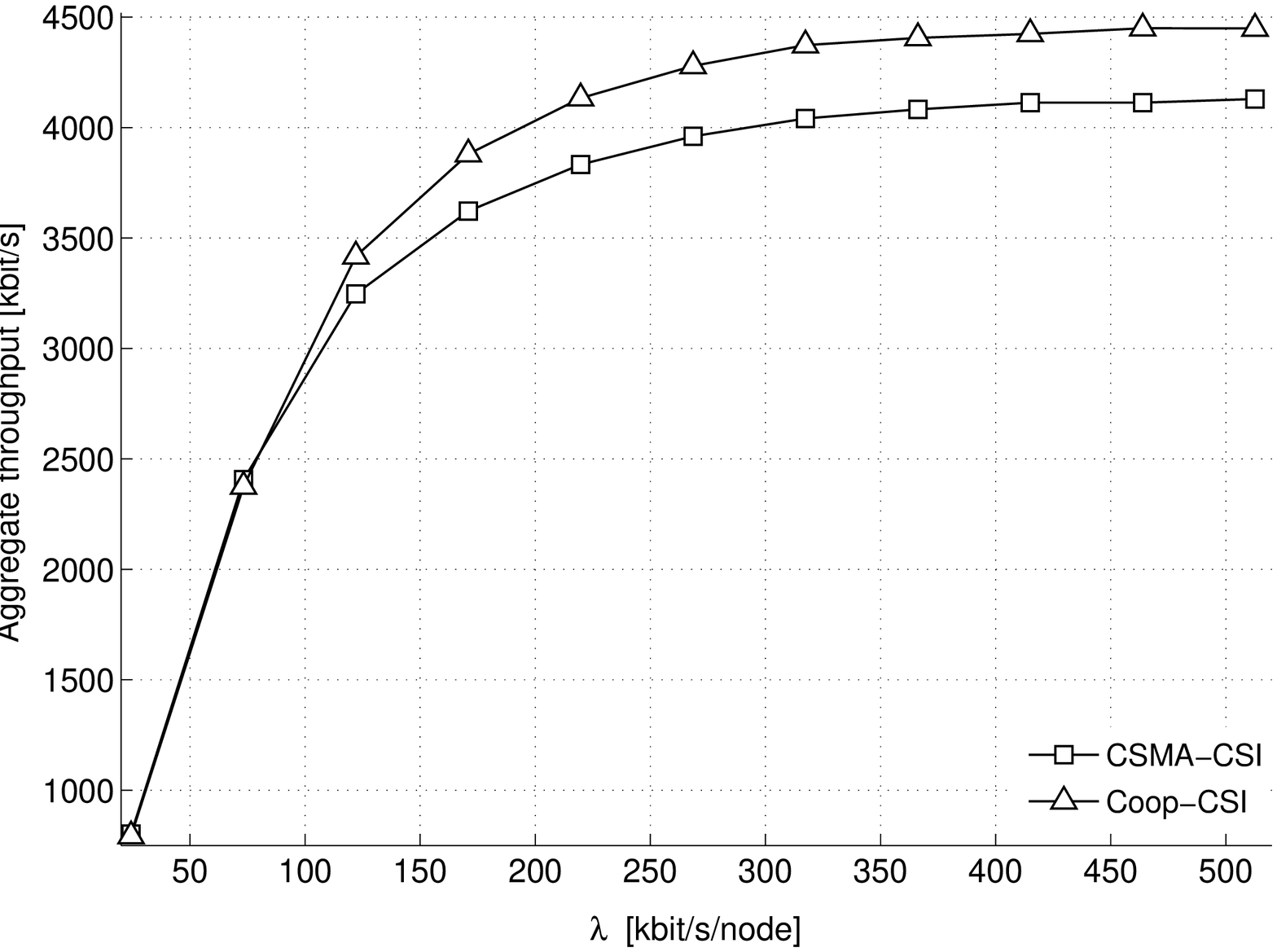}
\caption{Aggregate throughput vs nominal load.}
\label{fig:preSims_tru}
\end{figure}

As a general observation, our simulations have shown how the remarkable performance gains pointed out for relaying in simple topologies by the analytical framework of Section~\ref{sec:preAllocAnalysis} (see Fig.~\ref{fig:preAnalysis_truputBiased}) plummet when the protocols are implemented in large-scale ad hoc networks. Such a behavior is evident in Fig.~\ref{fig:preSims_tru}, which reports the aggregate network throughput against the nominal load $\lambda$, highlighting how the improvements achieved by Coop-CSI over CSMA-CSI are curbed to 10\% at saturation in the considered scenarios. We remark that this result is of broad applicability and of particular relevance, since, as discussed, the proposed strategies represent a bound for the class of protocols that take advantage of channel state information to realize proactive cooperation.

\begin{figure}
\centering
\includegraphics[width=\figw]{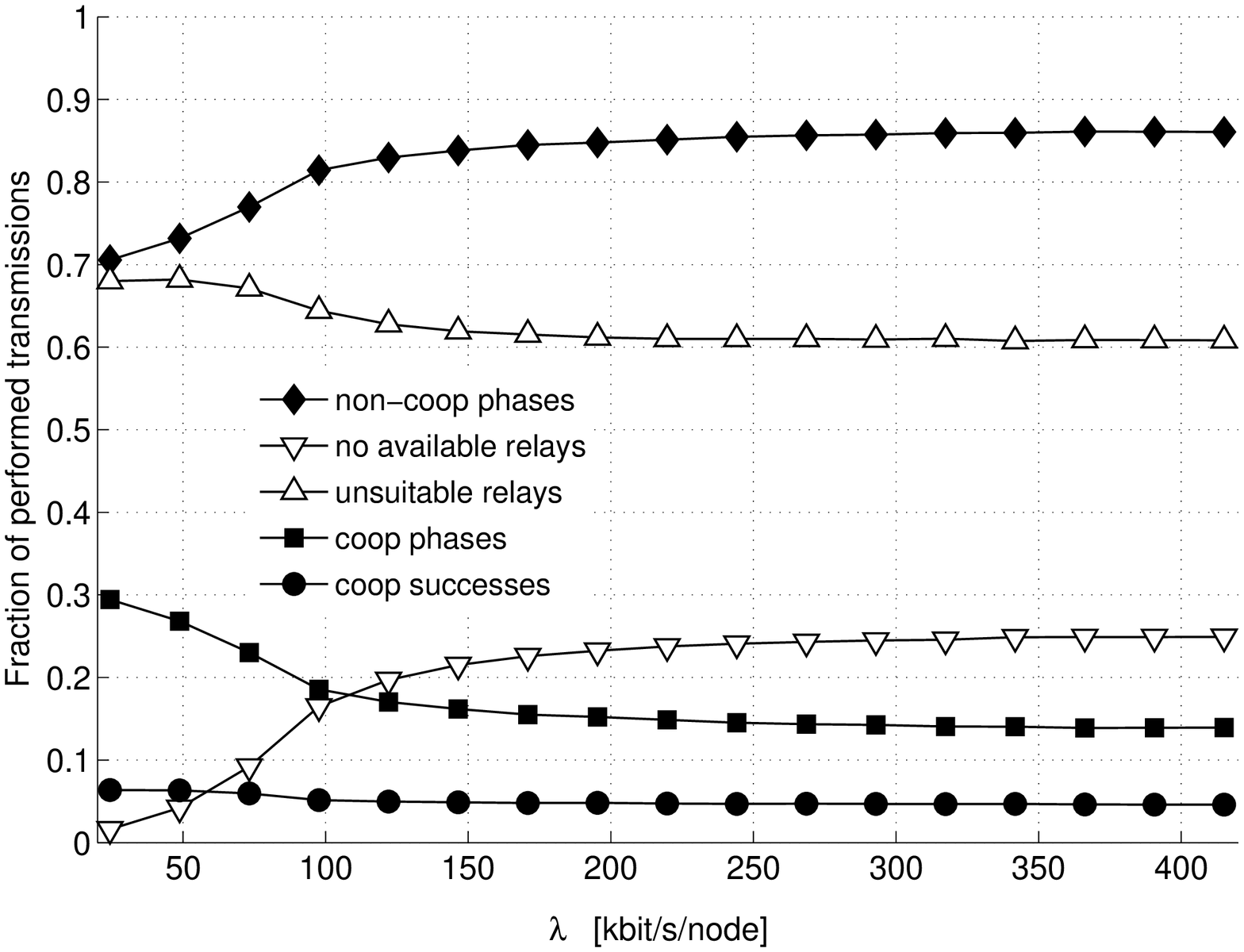}
\caption{Fraction of performed cooperative phases over all transmissions. Curves with empty markers indicate the reasons for not performing a cooperative transmission.}
\label{fig:preSims_coopImpact}
\end{figure}

\begin{figure}
\centering
\includegraphics[width=\figw]{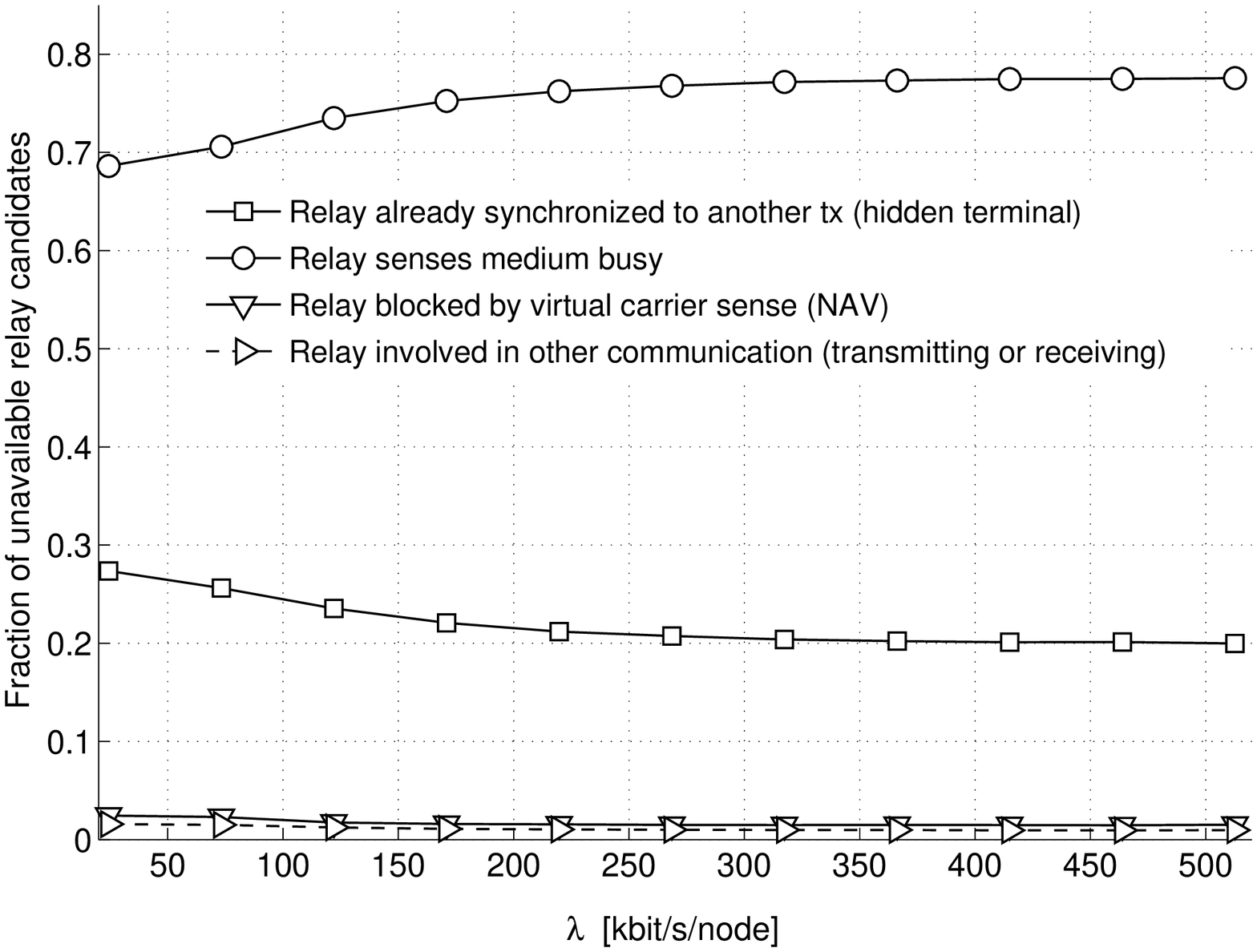}
\caption{Impact of the reasons that may lead a relay candidate to be unavailable for cooperation regardless of the channel quality it exhibits towards the destination.}
\label{fig:preSims_unavRelays}
\end{figure}

A first important insight on the issues that thwart the effectiveness of relaying is provided in Fig.~\ref{fig:preSims_coopImpact}. The plot highlights that Coop-CSI resorts to a two-hop link in less than 15\% of the initiated communications, whereas non-cooperative transmissions take place in almost 90\% of the cases. Recalling the discusssion of Section~\ref{sec:preAllocDescription}, packets are directly delivered over the source-destination channel if either i) no node is available for relaying, i.e., not involved in other ongoing activities and sensing an idle medium, or ii) some candidates are present but none of them is able to reduce the overall duration of the data transfer by cooperating. The impact of these factors is reported by the white-marked lines in the figure, which sum up to the \emph{non-coop phases} curve. As expected, the higher the traffic injected in the network, the lower the probability of finding terminals which are idle and allowed to access the channel due to the harsher interference level (\emph{no avail. relays}). Fig.~\ref{fig:preSims_unavRelays} delves into the reasons that lead to such a relay unavailability, showing how in the vast majority of cases cooperators are blocked by the power level sensed on the medium ($\sim$ 80\%) and, less prominently, by the hidden terminal problem ($\sim$ 20\%), while the impact of virtual carrier sense\footnote{Packet headers contain information on the duration of the communication they initiate. Therefore, nodes that decode a header not addressed to them also become aware that the channel will be busy for a given time, and suitably update their Network Allocation Vector (NAV) \cite{802.11}, deferring any channel access for the reserved period regardless of the power level perceived on the medium.} and of terminals that are already transmitting or receiving another packet is negligible. This result further supports the intuition that the low efficiency of reactive cooperation, discussed in the companion paper, is not related specifically to the protocol that we propose, but rather arises from the intrinsic nature of carrier sense multiple access.

Going back to the discussion of Fig.~\ref{fig:preSims_coopImpact}, the plot suggests that the main obstacle to the establishment of two-hop links is by far the lack of nodes that, while sensing a power level on the medium below the carrier sense threshold, are also capable of shortening transmission times (\emph{unsuitable relays}), responsible for more than 70\% of the non-relayed phases at saturation.
It is thus apparent that a first and major limitation to the effectiveness of Coop-CSI stems from the reduced share of cooperative links it triggers. On the other hand, Fig.~\ref{fig:preSims_coopImpact} also emphasizes that, even when performed, relayed transmissions are not efficient, with a success rate lower than 50\% (\emph{coop successes}).

\begin{figure}
\centering
\includegraphics[width=\figw]{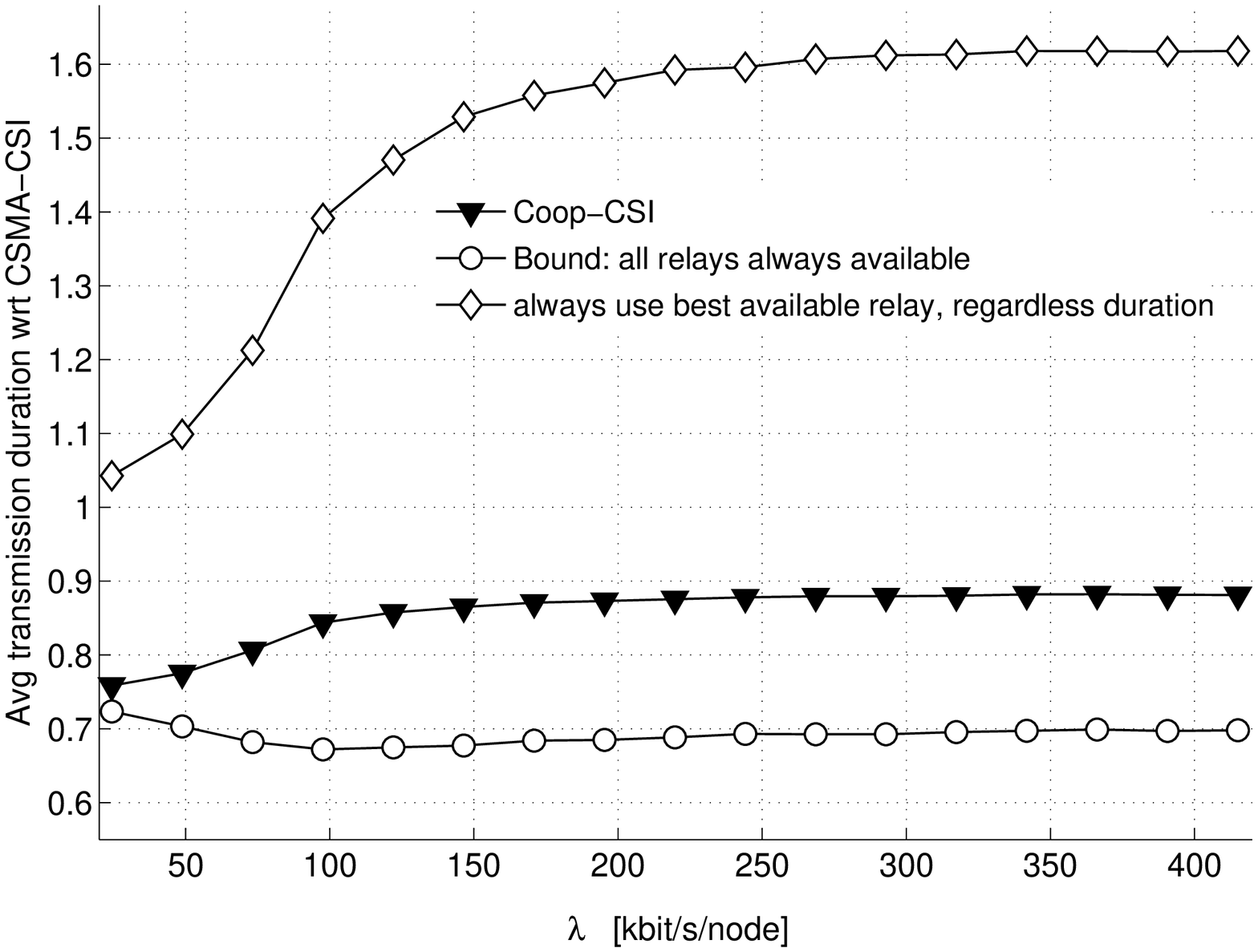}
\caption{Average duration of a transmission for different cooperative schemes normalized to the average transmission duration for CSMA-CSI.}
\label{fig:preSims_durationGain}
\end{figure}

The former issue is further investigated in Fig.~\ref{fig:preSims_durationGain}. Here, the black-marked line describes the ratio of the average duration for a complete communication that characterizes Coop-CSI (over either a direct or a split link) to the same quantity achieved with the plain CSMA-CSI strategy, and points out again the limited improvements discussed earlier in terms of aggregate throughput. Conversely, the white circle-marked curve shows the transmission duration gain that would be secured if all the neighbors of a source node were always available for cooperating, i.e., if the protocol could always count also on terminals that are actually prevented from relaying because of the power level sensed on the medium. The reported trend confirms the inferences prompted by our analytical discussion, e.g., Figs.~\ref{fig:preAnalysis_molare} and \ref{fig:preAnalysis_truputBiased}, highlighting how even in large-scale distributed networks proactive cooperation would lead to important performance gains (up to 30\%) if relays did not have to obey CSMA. Such a potential, however, is intrinsically encumbered by the medium acces policy, as nodes in the most favorable positions for cooperating, i.e., those that offer some advancement to the destination, are typically located at the periphery of the spatial region reserved by the carrier sense mechanism for the source's transmission and thus also experience a non-negligible chance of sensing a busy channel. This fundamental  discrepancy is further supported by the white diamond-marked curve in Fig.~\ref{fig:preSims_durationGain}, that plots the average transmission duration for Coop-CSI if the protocol always resorted to cooperation by selecting the best available relay candidate without considering the possibility of employing a direct link to the destination, i.e., applying (\ref{eq:minT}) without the $T_{s,d}$ term, normalized to the duration of a communication in CSMA-CSI. Once more it is manifest that terminals allowed by the medium access control to transmit trigger two-hop links with much poorer performance compared to the basic solution. The result, then, clearly explains the large share of untapped cooperative phases that characterizes Coop-CSI as discussed in Fig.~\ref{fig:preSims_coopImpact}. In this perspective it is also important to remark how these issues would not be mitigated by more aggressive medium access approaches, i.e., allowing relays to access the medium even if the power level they perceive is above $\Lambda$, as they inherently depend on the relationship between CSMA and cooperation. Dedicated simulations, whose outcome is not reported here due to space constraints, have shown that setting a less stringent carrier sense threshold $\Lambda_{rel}$ for cooperating terminals, i.e., $\Lambda_{rel} > \Lambda$, does not lead to an overall performance improvement. Indeed, the higher number of cooperative phases triggered by larger values of $\Lambda_{rel}$ is outweighed by their reduced effectiveness resulting from the increased aggregate interference level they generate. A more in-depth discussion of these tradeoffs can be found in \cite{part1}.

\begin{figure}
\centering
\includegraphics[width=\figw]{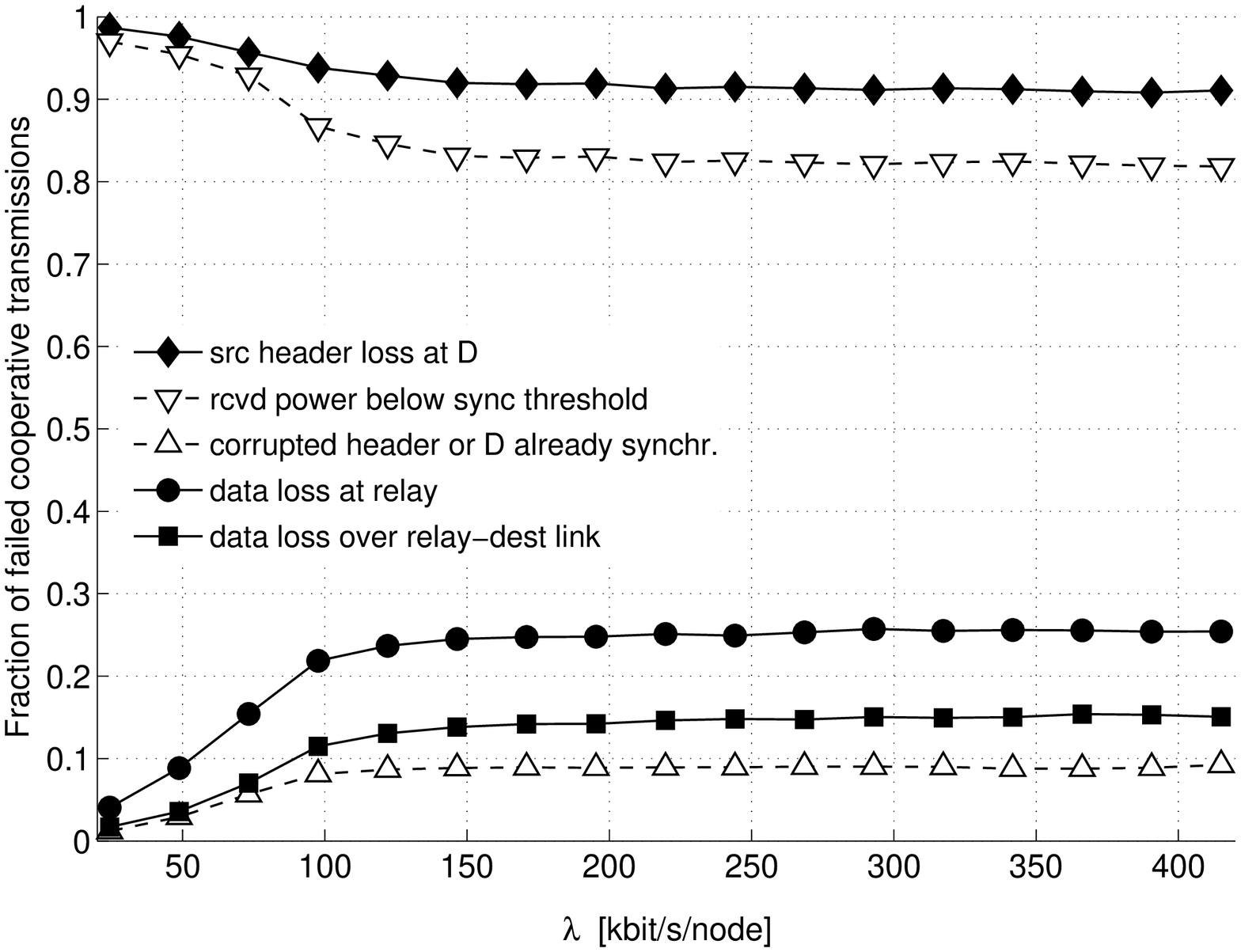}
\caption{Reasons that lead to the failure of cooperative transmissions.}
\label{fig:preSims_failures}
\end{figure}

Fig.~\ref{fig:preSims_failures} focuses on the reasons that may cause a cooperative link to fail once it has been triggered. In order for a two-hop communication involving terminals $\bm S$, $\bm C$ and $\bm D$ to be successful, three conditions have to be met: i) the data unit has to be retrieved at $\bm C$; ii) the destination has to decode the header of the packet sent over the $\bm S$-$\bm C$ link, so as to cache a corrupted version of the payload for later combining; and iii) the redundancy sent by the relay has to enable data decoding at $\bm D$. By virtue of the rate adaptation policy of Coop-CSI as well as the protection offered by the margin coefficient $\varepsilon$, the impact of factors ii) and iii) is rather limited (\emph{data loss at relay} and \emph{data loss over C-D link} curves). Conversely, the plot shows that the vast majority of cooperative phases fail since the destination does not retrieve the header of the payload sent by the source. In turn, such header losses may occur either because $\bm D$ does not synchronize to the incoming packet, or because the decoding does not succeed due to channel impairments, with the former reason possibly induced by the hidden terminal problem or by too low a received power level.\footnote{If the power level of an incoming packet is below a synchronization threshold, see Tab.~\ref{tab:parameters}, a node regards it as noise and does not even try to synchronize to it.} A remarkable conclusion of the companion paper \cite{part1} is that the foremost reason for the destination to miss a header addressed to it when \emph{reactive} cooperation is implemented is that the node is already synchronized to another ongoing transmission. Instead, Fig.~\ref{fig:preSims_failures} points out an insufficient received power as the key factor for cooperation not to be successful if \emph{proactive} schemes are employed (dashed white-marked curves). This difference can be explained in light of the spatial distribution for relay nodes induced by CSMA. As discussed, terminals available for proactive cooperation tend to be located with high probability in positions that do not allow particularly fast two-hop transmissions. Therefore, when a split link is actually triggered by Coop-CSI, the direct connection between source and destination is likely to offer extremely poor performance, in particular because affected by a deep fade. Such a condition, however, also weakens the useful power perceived at the addressee, hampering the header decoding and thus the effectiveness of relayed phases.

\begin{figure}
\centering
\includegraphics[width=\figw]{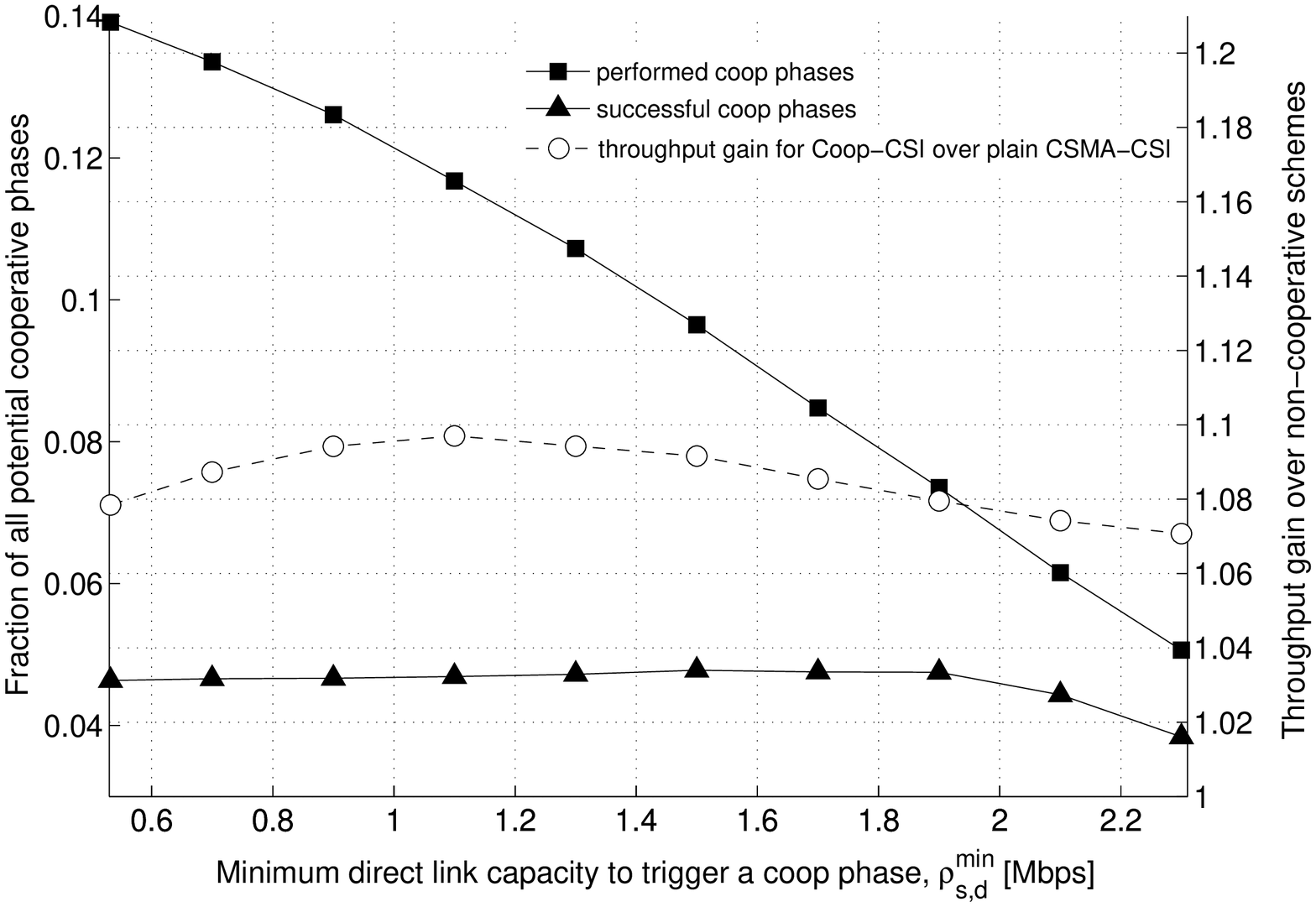}
\caption{Impact of cooperative phases and aggregate throughput gain with respect to CSMA-CSI when different thresholds of the minimum source-dest link capacity are considered for triggering cooperation. Left axis, solid lines: fraction of performed (squared markers) and successful cooperative phases (triangle-markers); right axis, white markers: throughput gain of Coop-CSI over CSMA-CSI, i.e., ratio of the aggregate throughputs achieved by the two protocols.}
\label{fig:rcSims_truVsMinRate}
\end{figure}

Following this argument, one could think of letting cooperation take place only when a minimum channel quality over the $\bm S$-$\bm D$ link is guaranteed. Nonetheless, while this approach would certainly increase the success probability of two-hop links, it may also reduce the already small share of cooperative communications that are performed. This tradeoff is apparent in Fig.~\ref{fig:rcSims_truVsMinRate}, which depicts the behavior of the protocol when such a mechanism is implemented. According to this modified version of Coop-CSI, a source node checks for potential cooperators only if the minimum sustainable rate $\rho_{s,d}$ with its destination is above a threshold $\rho_{s,d}^{min}$, reported on the $x$ axis of the plot. If so, the procedures described in Section~\ref{sec:Coop-CSI} take place as usual. Otherwise, $\bm S$ resorts to a direct link with $\bm D$ or just drops the packet if $\rho_{s,d} < \rho_{min}$.\footnote{Notice that the basic version of the protocol simply sets $\rho_{s,d}^{min} = \rho_{min} = 0.9\,$Mbps, as per Tab.~\ref{tab:parameters}.} The figure highlights how an improvement in terms of reliability for proactive cooperation, i.e., higher values of $\rho_{s,d}^{min}$, comes at the expense of a drastic fall of the number of relaying phases. The outcome of such  a compromise at a system level is reported by the white-marked line, whose values have to be referred to the right $y$ axis. The throughput gain over CSMA-CSI does not vary significantly, proving once more how the poor performance of proactive relaying in large-scale ad hoc networks considered here is intrisically related to the effects of the carrier sense-based medium access policy and not a consequence of the specific protocol implementations proposed in this work.

%% file: conclusions.tex
\section{Conclusions} \label{sec:conclusions}

This work is the second of a two-part series of papers addressing the effectiveness of cooperative communications in non-centralized ad hoc wireless networks with carrier sense-based channel contention. Part I investigated \emph{reactive} cooperative approaches, in which relaying is triggered in response to packet failures. This paper, instead, discussed \emph{proactive} solutions, in which source and relays take advantage of CSI to preemptively coordinate and maximize the sustainable data rate over a source-destination link. 

We showed through analysis and simulation that the contention mechanism, as well as the presence of interfering nodes, significantly diminishes the effectiveness of such a cooperative approach and bounds the gains predicted from a theoretical perspective in toy topologies with idealized medium access. First, the carrier sense access policy biases the spatial distribution of nodes available to offer cooperative support to a source in delivering its traffic, reducing the probability that they lie in the spatial region that would maximize the performance gains. Secondly, the completely uncoordinated nature of medium access in the networks under consideration leads to a highly dynamic interference level, which, in turn, alters the boundary conditions used by nodes to preemptively set up cooperative links, and thus jeopardizes the overall reliability of the strategy. Finally, the impact of several practical issues, such as the synchronization and the hidden terminal problem, has been studied.